# Tunable ultraviolet dispersive-wave emission driven directly by 40-fs Ti: sapphire laser pulses in hollow capillary fiber

TIANDAO CHEN,[1,3,4] ZHIYUAN HUANG,[1,2,*] JINYU PAN,[1] DONGHAN LIU,[1,2,3] YINUO ZHAO,[1,3] WENBIN HE,[2] JIAPENG HUANG,[2] XIN JIANG,[2] MENG PANG,[1,2,5] YUXIN LENG,[1,6] AND RUXIN LI[1,4]

[1]*State Key Laboratory of High Field Laser Physics and CAS Center for Excellence in Ultra-intense Laser Science, Shanghai Institute of Optics and Fine Mechanics (SIOM), Chinese Academy of Sciences (CAS), Shanghai 201800, China*
[2]*Russell Centre for Advanced Lightwave Science, Shanghai Institute of Optics and Fine Mechanics and Hangzhou Institute of Optics and Fine Mechanics, Hangzhou, 311421, China*
[3]*Center of Materials Science and Optoelectronics Engineering, University of Chinese Academy of Sciences, Beijing 100049, China*
[4]*Zhangjiang Laboratory, Shanghai 201210, China*
[5]*e-mail: pangmeng@siom.ac.cn*
[6]*e-mail: lengyuxin@siom.ac.cn*
*\*Corresponding author: huangzhiyuan@siom.ac.cn*

**We demonstrate that by using 1-m-long gas-filled hollow capillary fiber (HCF) with a core diameter of 100 μm, tunable ultraviolet (UV) dispersive-wave (DW) pulses can be generated in a compact, single-stage set-up driven directly by 40-fs Ti: sapphire laser pulses. By adjusting the gas type and pressure inside the HCF, the central wavelength of the UV DW can be continuously tuned from 185 nm to ~450 nm. In the experiment, we found that for longer-wavelength (from ~320 to ~450 nm) DW generation, Raman-active gas filled in the HCF can efficiently suppress the pulse splitting effect of the high-order soliton due to the Raman-induced pulse energy dissipation, leading to the high-quality DW generation at these wavelengths with smooth, single-peak spectra. These results provide some useful insights for designing compact, wavelength-tunable ultrafast UV light sources with microjoule-level pulse energies.**

Few-femtosecond optical pulses with continuous wavelength tunability from ultraviolet (UV) to visible region are crucial for a few applications, such as transient absorption spectroscopy [1-5] and ultrafast photochemical dynamics [3-8]. Phase-matched dispersive-wave (DW) emission in gas-filled hollow capillary fibers (HCFs), as an important method of generating microjoule-level wavelength-tunable ultrashort pulses in the UV and visible regions, has attracted a wide range of research interest in recent several years [9-12]. Compared with traditional ultrafast frequency-conversion processes such as optical parametric amplification or harmonic generation, phase-matched DW emission can generate ultrashort laser pulses in short-wavelength region with high conversion efficiency and broadband wavelength tunability. Through adjusting the gas type and pressure filled into the hollow-core waveguide, the phase-matched DW wavelength can be continuously tuned from visible to vacuum UV [9-12].

In 2019, Travers *et al*. demonstrated that 3-m-long gas-filled HCF can be used to generate continuously-tunable, high-energy DW pulses at vacuum- and deep-UV wavelengths [9]. Recently, Brahms *et al*. used the scaling rule to set up a more compact DW-emission system, which can be used to generate tunable ultrashort deep-UV pulses [10]. Through using smaller-core HCF and shorter pump pulses, they demonstrated that the HCF length used in the system can be shortened to ~15 cm [10]. However, in previous studies few-cycle pulses were normally required for driving the high-efficient DW emission process, leading to two-stage gas-filled HCF systems in all these experiments [9-12]. While the first stage is for pulse compression to ~10-fs or even shorter duration, the compressed pulse was then used in the second stage of HCF system for high-efficient DW emission.

In this work, we demonstrated that microjoule-level DW pulses with broadband wavelength tunability can be generated from a single-stage HCF system, driving directly by commercial Ti: sapphire ultrafast laser pulses with 40 fs durations. In the experiment, through carefully selecting the gas type and pressure, and at the same time adjusting the pulse energy launched into the hollow-core waveguide, we obtained in the 100-μm-diameter HCF a fast compression process of the driving pulse with a high (>10) soliton order from 40 fs to ~2 fs, giving rise to the high-efficient emission of DWs with smooth optical spectra. Moreover, we demonstrated experimentally and numerically that at the condition of 40-fs driving pulse, $N_2$ (Raman-active) gas can be used to suppress the pulse splitting phenomenon in the high-order soliton propagation. Thanks to the effect of Raman-induced pulse energy dissipation, the appearance of multi-peak structure on the DW spectrum can be avoided, leading to the generation of smooth, single-peak spectra over the whole wavelength tuning range from 185 nm to ~450 nm.

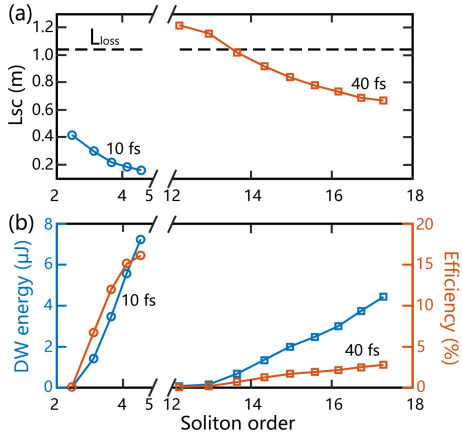

Fig. 1. (a) Self-compression length ($L_{SC}$, extracted from simulation results) scaling in 5-bar Ne-filled, 100-μm-diameter HCF, for 10 fs (blue line) and 40 fs (orange line) pulses with a central wavelength of 800 nm. $L_{loss}$, 10-dB power loss length. (b) DW energy (blue line) and pump-to-DW efficiency (orange line) versus soliton order for different pulse duration.

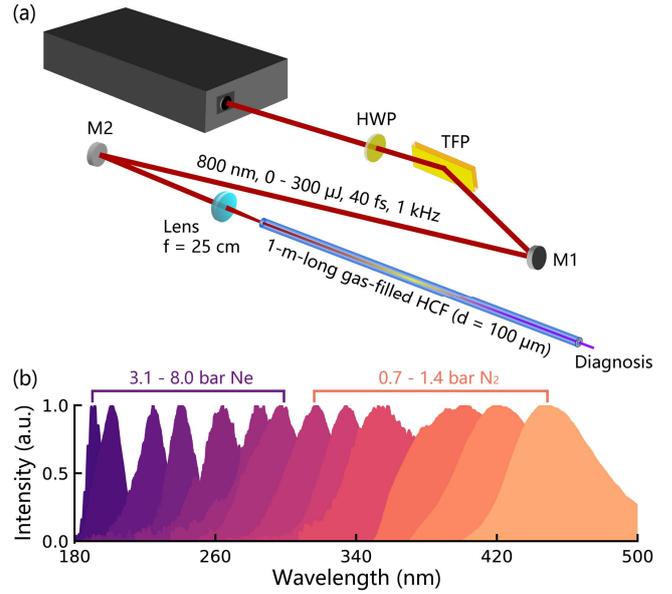

Fig. 2. (a) The sketch of the experiment set-up. HWP, half-wave plate; TFP, thin-film polarizer; M1-M2, silver mirror; HCF, hollow capillary fiber. (b) DW spectra at different emission wavelengths, measured using a fiber-coupled calibrated spectrometer (Ocean Insight Maya 2000Pro) and an integrating sphere.

The difficulty for generating high-efficient DW with driving pulses of higher temporal widths, lies in the relative scales between the effective length of the HCF ($L_{loss}$) and the self-compression length of the driving high-order soliton ($L_{SC}$) [9]. For example, the HCF with a core diameter of 100 μm has an effective HCF length of ~1 m at 800 nm (the driving pulse wavelength) for a HCF loss value of 10 dB, as illustrated in Fig. 1(a). The high-efficient DW emission in such a HCF system requires that the $L_{SC}$ of the driving pulse is shorter than $L_{loss}$, since the DW emission phenomenon is the consequence of the optical spectral broadening accompanied with the soliton self-compression [13]. As illustrated in Fig. 1(a), while at a fixed pulse width, $L_{SC}$ decreases as the soliton order ($N$) increases due to enhanced optical nonlinearity [9,13], for a driving pulse with a higher temporal width, a higher soliton order is needed in the same HCF system to ensure that $L_{SC}$ is shorter than $L_{loss}$.

This prediction is verified in our numerical simulations. Using the single-mode unidirectional pulse propagation equation [14,15], we performed numerically pulse propagation simulations in 1-m-long, 100-μm-diameter HCF filled with 5-bar Ne gas (corresponding to a phase-matched DW wavelength of ~240 nm), and the results are illustrated in Fig. 1(b). It can be found that for 10-fs pulse width, a pump pulse with a soliton order of 2.5 to 4.6 can efficiently drive the DW emission process, leading to a high frequency-conversion efficiency up to 15%. In contrast, for the 40-fs pulse-width case, the soliton order needed for driving DW emission is as high as ~14. Such a high soliton-order value largely shortens, in one side, the self-compression length of the pulse in the HCF, in the other side, the higher soliton order gives rise to a higher pulse compression ratio, strongly broadening the pulse optical spectrum which covers the phase-matched DW wavelength as the prerequisite of the high-efficiency DW emission [13].

The compact single-stage set-up used in our experiments is illustrated in Fig. 2(a). The driving pulses of the system comes directly from a commercial Ti: sapphire femtosecond laser (Legend Elite, Coherence) which can deliver mJ-level 800-nm pulses with 40 fs pulse width. After passing through a half-wave plate and a thin-film polarizer (working as a tunable attenuator), the driving pulses were launched into the 100-μm-diameter HCF with a fiber length of 1 m through an optical lens (25 cm focal length), giving a measured coupling efficiency of ~75%. The launched pulse energy could be varied through adjusting the half-wave plate.

When the HCF was filled with 5-bar Ne, we measured the evolution of the output spectrum from the HCF system as the function of input pump energy, see Fig. 3(a). As the input pump energy was increased to ~110 μJ (corresponding to a soliton order value of ~14), clean DW spectrum at ~240 nm was observed on the output spectrum (see Fig. 3(a)). At an input pump energy of 157 μJ, the generated UV DW energy was estimated to be 1.4 μJ, corresponding to a conversion efficiency of 0.89%. These results exhibit striking agreement with the theoretical prediction demonstrated in Fig. 1. For a comparison, simulation results of the output spectral evolution are illustrated in Fig. 3(b). In the experiment, we gradually adjusted the Ne-gas pressure inside the HCF from 3.1 bar to 8.0 bar, by doing this the phase-matched DW wavelength could be shifted from 185 nm to 296 nm [13,15-18]. During this tuning process, we adjusted accordingly the input pump energy so as to ensure the soliton order value of the driving pulse to be ~17. Then, we obtained continuously-wavelength tuning (185 – 296 nm) of the DW emission, as illustrated in Fig. 2(b). The DW pulse energy was estimated to be 0.6 μJ – 1.8 μJ over the whole wavelength-tuning range.

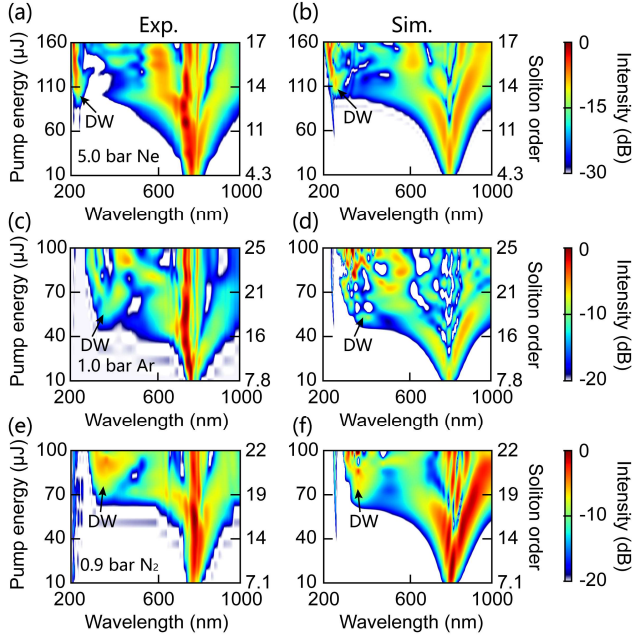

Fig. 3. Experimental (a, c, e) and simulated (b, d, f) spectral evolutions of the pulse at the output end of HCF as a function of pump energy. (a, b), (c, d), and (e, f) correspond to 5 bar Ne, and 1 bar Ar, and 0.9 bar $N_2$ respectively. DW, dispersive wave.

Pushing the DW emission to longer wavelengths (>300 nm) in this Ne-gas system, requires even higher gas pressure. An alternative scheme for generating longer-wavelength DW is to change the gas type filled into the HCF from Ne to Ar [18], which can result in the same phase-matched wavelength at a relatively-lower pressure. However, in the experiment we found that in an Ar-filled HCF system, the generated DW spectra at longer wavelengths exhibit normally broadband features with multi-peak spectral structures. One typical example is illustrated in Figs. 3(c) and 3(d), when the HCF was filled with 1-bar Ar gas (corresponding to a phase-matched DW wavelength of ~370 nm), the DW emission can be obtained in the experiment at input pulse energies of >55 μJ, corresponding to soliton order values of >18, see Fig. 3(c). As the input power energy increased, the generated DW spectrum became broader with multi-peak features, which could be perfectly retrieved in the numerical simulations, see Fig. 3(d). This obvious distortion of DW spectrum can be understood as: in order to obtain phase-matched DW wavelength at longer wavelengths, the zero-dispersion wavelength of the HCF waveguide is shifted to even longer wavelengths. As the zero-dispersion wavelength approaches to the pump wavelength of the driving pulse (800 nm in the experiment), the symmetry of high-order soliton propagation would be broken up [12,13], leading to strong high-order-soliton splitting effects. At a soliton order value as high as >18, the soliton splitting could result in the generation of multiple sub-pulses, in the temporal domain, with different pulse energies and group velocities, co-propagating with the generated UV DW. Strong cross-phase modulations would cause the distortion of the DW spectrum [12], as illustrated in Figs. 3(c) and 3(d).

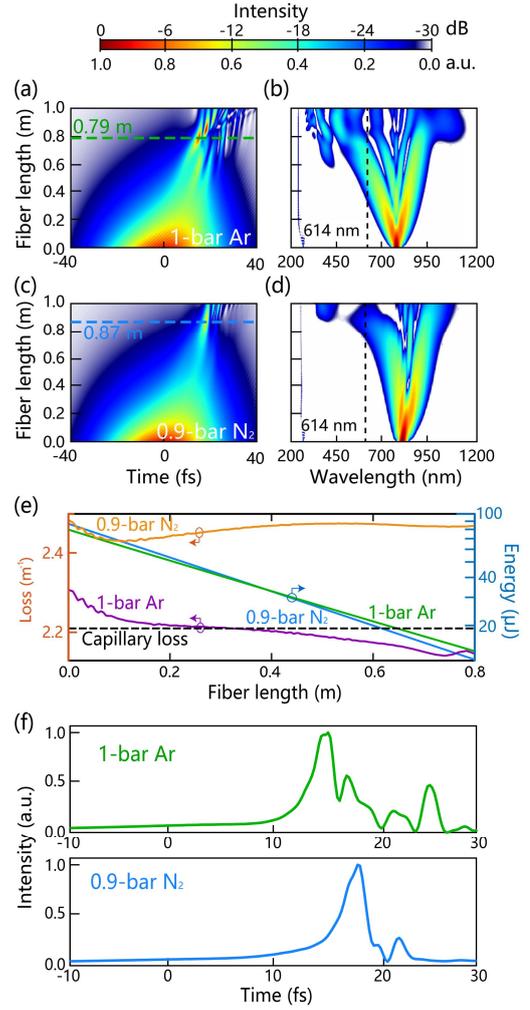

Fig. 4. Simulated temporal (a, c) and spectral (b, d) evolutions of the pulse as a function of fiber length. (a, b) and (c, d) correspond to 1 bar Ar and 0.9 bar $N_2$, respectively. Black dashed lines in (b) and (d) indicate the zero-dispersion wavelength of 614 nm. (e) Loss (orange and purple lines) and energy (blue and green lines) profiles of the pulse as a function of fiber length. Black dashed line indicates the capillary loss for the pulse centered at 800 nm. (f) Temporal profiles of the pulse at self-compression points marked as green and blue dashed lines in (a) and (c).

In the experiment, we found that such spectral distortion of DW emission could be largely mitigated through introducing Raman-active gas. As illustrated in Figs. 3(e) and 3(f), when 0.9-bar $N_2$ gas was used in the HCF core, relatively-smooth DW spectra could be obtained at high soliton orders, compared with the results shown in Fig. 3(c). This phenomenon can be better understood through simulating the pulse propagation process along the HCF as illustrated in Figs. 4(c)-4(e), it can be found that in Raman-active ($N_2$) gas, the excitation of Raman molecular vibrations results in some extra loss [19,20] to the driving pulse (see Fig. 4(e)), decreasing gradually the soliton order over the HCF length. The soliton splitting effect near and after the pulse self-compression point can therefore be suppressed due to the relatively-lower soliton order.

The nonlinear propagation process of 80 μJ, 40-fs pulse (soliton order of ~22) in the 1-bar-Ar-filled, 1-m-long HCF, is illustrated in Figs. 4(a) and 4(b). The simulation results exhibit that the self-compression point in the HCF is at ~0.79 m, and soliton splitting effects are observed near and after the self-compression point, see Fig. 4(f) for the temporal pulse profile at exactly the compression point. At the similar driving pulse parameters (86 μJ, 40 fs), the use of 0.9-bar $N_2$ gas (soliton order of ~21) in the HCF can lead to a dramatic change of both the spectral and temporal evolutions of the pulse, especially near and after the self-compression point ($L_{SC}$ = 0.87 m), see Figs. 4(c) and 4(d). It can be found that compared with the Ar-gas case (see Fig. 4(a)), the extra loss due to Raman effects [19,20] decreases the soliton order along the HCF length, suppressing the soliton splitting effect [13,17]. Such suppression of soliton splitting is further verified by the temporal pulse profile at the compression point, see Fig. 4(f). This clean pulse profile ensures the generation of the smooth DW spectrum without distortion, through avoiding the cross-phase modulation effects from the sub-pulses. See some more detailed results and discussions in the Supplement 1. Using the $N_2$-filled HCF system, μJ-level UV DW pulses with tunable wavelengths from ~320 nm to ~450 nm and relatively-smooth spectral profiles, can be obtained through adjusting the $N_2$ gas pressure from 0.7 bar to 1.4 bar, giving rise to frequency conversion efficiencies of 0.63% to 1.25%, see some detailed data in Fig. 2(b) and Supplement 1.

In conclusion, we demonstrated both experimentally and numerically that microjoule-level, wavelength-tunable DW pulses can be generated in a single-stage gas-filled HCF system, pumped directly by a commercial 40-fs Ti: sapphire laser. We found that through increasing the soliton order of the driving pulse, the self-compression length of the pump pulse in the HCF can be efficiently decreased to shorter than the effective length of the high-loss HCF, leading to the high-efficiency generation of the UV DW. Moreover, in the experiment Raman-active ($N_2$) gas can be used to suppress the splitting effects of the high-order soliton near and after the self-compression point, avoiding the distortion of the DW spectra at longer wavelengths. The results demonstrated here point out a compact HCF set-up for generating high-energy ultrafast laser pulses with broadband wavelength tunability in UV and visible regions.

**Funding.** National Natural Science Foundation of China (62205353, 62275254, 61925507 and 12388102), the Strategic Priority Research Program of the Chinese Academy of Science (XDB0650000), the National Postdoctoral Program for Innovative Talents (BX2021328), the China Postdoctoral Science Foundation (2021M703325), the Shanghai Science and Technology Plan Project Funding (23JC1410100), the Shanghai Science and Technology Innovation Action Plan (21ZR1482700), National Key R&D Program of China (2022YFA1604401), Shanghai Science and Technology Committee Program (22DZ1100300, 22560780100 and 23560750200).

# Tunable ultraviolet dispersive-wave emission driven directly by 40-fs Ti: sapphire laser pulses in hollow capillary fiber: supplement

TIANDAO CHEN,[1,3,4] ZHIYUAN HUANG,[1,2,*] JINYU PAN,[1] DONGHAN LIU,[1,2,3] YINUO ZHAO,[1,3] WENBIN HE,[2] JIAPENG HUANG,[2] XIN JIANG,[2] MENG PANG,[1,2,5] YUXIN LENG,[1,6] AND RUXIN LI,[1,4]

[1]*State Key Laboratory of High Field Laser Physics and CAS Center for Excellence in Ultra-intense Laser Science, Shanghai Institute of Optics and Fine Mechanics (SIOM), Chinese Academy of Sciences (CAS), Shanghai 201800, China*
[2]*Russell Centre for Advanced Lightwave Science, Shanghai Institute of Optics and Fine Mechanics and Hangzhou Institute of Optics and Fine Mechanics, Hangzhou, 311421, China*
[3]*Center of Materials Science and Optoelectronics Engineering, University of Chinese Academy of Sciences, Beijing 100049, China*
[4]*Zhangjiang Laboratory, Shanghai 201210, China*
[5]*e-mail: pangmeng@siom.ac.cn*
[6]*e-mail: lengyuxin@siom.ac.cn*
*\*Corresponding author: huangzhiyuan@siom.ac.cn*

# Tunable ultraviolet dispersive-wave emission driven directly by 40-fs Ti: sapphire laser pulses in hollow capillary fiber: supplement document

## 1. Influences of Raman and ionization effects

We performed numerical simulations using the unidirectional pulse propagation equation to investigate the influences of Raman effect and gas ionization on dispersive-wave (DW) emission in $N_2$-filled hollow capillary fiber (HCF). As illustrated in Figs. S1(a) and S1(b), when the Raman effect is turned off, soliton splitting effects are observed near and after the self-compression point, which is very similar to Figs. 4(a) and 4(b) in the main text, confirming that the Raman effect is the main factor suppressing soliton-splitting dynamics. In addition, we also found that the gas ionization has an impact on the DW emission process. When the gas ionization is not included in the simulation, we can observe the broadening of DW spectrum, see Figs. S1(c) and S1(d). This broadband DW generation can be understood as: when the loss introduced by gas ionization is not considered, the pulse propagating in the HCF has a higher soliton order. Such a high soliton order could trigger soliton splitting effect, resulting in the generation of multiple sub-pulses at the trailing edge of the pulse, see Fig. S1(c). The sub-pulses, overlapping and co-propagating with the DW pulse generated in the HCF, lead to the broadening of DW spectrum through cross-phase modulation [1], see Fig. S1(d).

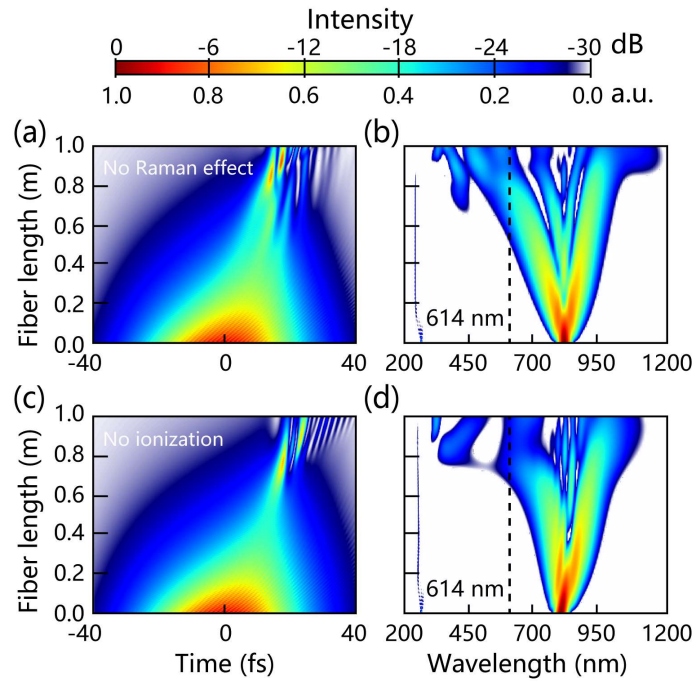

**Fig. S1.** Simulated temporal (a, c) and spectral (b, d) evolutions of the pulse propagating in the $N_2$-filled HCF with Raman effect turned off (a, b) and ionization turned off (c, d). All other simulation parameters are the same as Figs. 4(c) and 4(d) in the main text. Black dashed lines in (b) and (d) indicate the zero-dispersion wavelength of 614 nm.

Table S1. Parameters of tunable DW in Fig. 2(b).

| Gas type | Pressure (bar) | Pump energy[a] (μJ) | Soliton order | DW wavelength[b] (nm) | DW energy (μJ) | Efficiency[c] (%) |
|---|---|---|---|---|---|---|
| Ne | 3.1 | 240 | 16 | 185 | 0.6 | 0.25 |
|    | 3.8 | 183 | 16 | 204 | 1.0 | 0.55 |
|    | 5.0 | 157 | 17 | 225 | 1.4 | 0.89 |
|    | 6.0 | 110 | 16 | 241 | 1.5 | 1.36 |
|    | 6.7 | 101 | 17 | 264 | 1.2 | 1.19 |
|    | 7.2 | 122 | 19 | 283 | 1.8 | 1.48 |
|    | 8.0 | 98  | 18 | 296 | 1.4 | 1.43 |
| $N_2$ | 0.6 | 127 | 19 | 318 | 0.9 | 0.71 |
|    | 0.7 | 112 | 20 | 334 | 0.7 | 0.63 |
|    | 0.8 | 103 | 21 | 355 | 1.0 | 0.97 |
|    | 1.0 | 88  | 22 | 402 | 1.1 | 1.25 |
|    | 1.2 | 78  | 25 | 420 | 1.4 | 1.79 |
|    | 1.3 | 75  | 26 | 447 | 1.2 | 1.60 |

[a] Estimated from low-energy coupling efficiency.

[b] Calculated based on the peak value of the DW spectrum.

[c] Estimated from low-energy coupling efficiency.